\documentstyle[prl,aps,psfig]{revtex}

\begin{document}

\twocolumn[\hsize\textwidth\columnwidth\hsize\csname@twocolumnfalse\endcsname

\title{The Schottky defect formation energy in MgO calculated by
diffusion Monte Carlo} 

\author{D. Alf\`{e}$^{1,2}$, and M. J. Gillan$^2$
\smallskip \\ $^1$Department of Earth Sciences, University College
London \\ Gower Street, London WC1E~6BT, UK
\smallskip \\ $^2$Department of Physics and Astronomy, University
College London \\ Gower Street, London WC1E~6BT, UK}

\maketitle
\draft

\begin{abstract}
The energetics of point defects in oxide materials plays a major role
in determining their high-temperature properties, but experimental
measurements are difficult, and calculations based on density
functional theory (DFT) are not necessarily reliable. We report
quantum Monte Carlo (QMC) calculations of the formation energy $E_{\rm
S}$ of Schottky defects in MgO, which demonstrate the feasibility of
using this approach to overcome the deficiencies of DFT. In order to
investigate system-size errors, we also report DFT calculations of
$E_{\rm S}$ on repeating cells of up to $\sim 1000$~atoms, which
indicate that QMC calculations on systems of only 54 atoms should
yield high precision. The DFT calculations also provide the relaxed
structures used in the variational and diffusion Monte Carlo
calculations. For MgO, we find $E_{\rm S}$ to be in close agreement
with results from DFT and from model interaction potentials, and
consistent with the scattered experimental values. The prospects for
applying the same approach to transition metal oxides such as FeO are
indicated.
\end{abstract}


]

The quantum Monte Carlo (QMC) technique~\cite{foulkes01} is important
in condensed matter science, because it is generally much more
accurate than density functional theory (DFT) and allows one to make
accurate predictions for problems where DFT
fails~\cite{leung99,hood03,healy01,filippi02,grossman95}. It is
usually competitive in accuracy with high-level quantum chemistry
methods, but it has the advantage of being practicable for large
systems containing hundreds of atoms. Recently, efforts have been made
to apply QMC to oxide systems, including transition metal
oxides~\cite{needs03,lee}, for which DFT gives poor predictions for
magnetic properties, phonon frequencies and other
properties~\cite{massidda99}. We have recently reported~\cite{alfe05}
a QMC study of bulk MgO for which we find excellent agreement with
experiments for the bulk lattice parameter and bulk modulus, provided
appropriate corrections are made. We report here what we believe to be
the first QMC calculation of an oxide lattice defect energy, namely
the Schottky formation energy $E_{\rm S}$ in MgO. Defect energies in oxide
materials are technologically important for applications
ranging from high-temperature superconductors to
radioactive waste disposal, but are also
very difficult to measure experimentally. We shall show that our
results for $E_{\rm S}$ in MgO are consistent with
available experimental data, as well as supporting earlier 
DFT predictions~\cite{devita92}.

The Schottky energy $E_S$ is the energy required to form a cation and
anion vacancy pair, and governs the thermal equilibrium concentration
of vacancies~\cite{stoneham01}. The calculation of $E_S$ in ionic
materials has a very long history, going back to the very early work
of Mott and Littleton~\cite{mott38,catlow82,harding90}. DFT
calculations on defect formation and migration energies in oxides
first became possible in the early 1990's, and are now routinely
performed. However, particularly in transition metal oxides, the
reliability of DFT calculations is questionable.

QMC calculations are usually performed in two
stages~\cite{foulkes01}. In the first, known as variational Monte
Carlo (VMC), a trial many-body wavefunction is constructed as a
product of a Slater determinant of single-electron orbitals and the
so-called Jastrow factor which explicitly accounts for electronic
correlation. Since VMC by itself is not usually accurate enough, the
second stage is to use the many electron wavefunction produced by VMC
in diffusion Monte Carlo (DMC), which improves the ground state
estimate by performing an evolution in imaginary time. This would
yield the exact ground-state energy but for the fact that the trial
wavefunction fixes the nodes of the many-body wavefunction. Because of
this, the DMC energy is an upper bound to the true ground-state
energy, but for systems with a large band gap the difference is
expected to be very small.

QMC calculations of defect energies in any material are challenging,
and their feasibility is not obvious, for several reasons. First, it
is not yet routinely possible to perform structural relaxation with
QMC, and relaxation around defects produces a very large energy
lowering in ionic materials~\cite{mott38,catlow82,harding90}. Second,
defect energies must be obtained as a difference of two large
energies, both of which suffer the statistical errors inevitable with
Monte Carlo methods. Third, there is a system size error associated
with the limited size of the periodically repeated cell used in
condensed-matter QMC calculations, and the difficulty of going to very
large cell sizes makes it difficult to assess the error. To overcome
these problems, we employ DFT calculations in tandem with QMC. In
particular, the relaxed ionic positions used for QMC calculations on
the defective crystals are taken from DFT calculations.

The overall strategy for our calculations is as follows.  The Schottky
energy $E_{\rm S}$ is defined to be the energy change when a Mg$^{2+}$
ion and an O$^{2-}$ ion are removed from the MgO perfect crystal, the
resulting defective crystal is allowed to relax, and the removed ions
are replaced to form new perfect crystal. The two vacancies formed in
this process are supposed to be very far apart, so that there is no
interaction between them. In calculating $E_{\rm S}$, either by DFT or
by QMC, it is preferable to perform calculations in which no more than
a single vacancy, either a Mg$^{2+}$ vacancy or an O$^{2-}$ vacancy,
is present. We avoid doing calculations with both vacancies present in
the same system, because with the cell sizes that can be achieved in
practice, such calculations would suffer from a large unwanted
interaction between the vacancies.  Denoting by $E_N ( \nu^+ , \nu^-
)$ the relaxed total energy of a crystal containing $N$ cation lattice
sites and $N$ anion lattice sites, and with $\nu^+$ cation vacancies
and $\nu^-$ anion vacancies, we therefore express the Schottky energy
as $E_{\rm S} = E_N ( 1 , 0 ) + E_N ( 0 , 1 ) - \left[ 2 ( N - 1 ) / N
\right] E_N ( 0 , 0 )$.

The energies $E_N ( 1 , 0 )$ and $E_N ( 0 , 1 )$ both refer to
periodic systems in which the supercell has a net charge.  To make
these energies well defined, we must assume that the systems have been
rendered electrically neutral by the introduction of a uniform
background charge, as described in earlier
papers~\cite{devita92,leslie85,makov95}. The effect of the background
charge densities is to make the zero-wavevector terms in the Coulomb
energy finite rather than infinite, but these charge densities do not
appear explicitly in any other way in either DFT or QMC.

In this way of doing the calculations, the interaction of the charged
vacancies with their periodic images gives finite-size corrections
$\Delta E$ to the total energy, which scale as $L^{-1}$, where the
length $L$ characterises the dimensions of the supercell. As is well
known~\cite{devita92,leslie85}, these corrections can be quite
accurately approximated by the formula $\Delta E \simeq \alpha q^2 /
\epsilon_0 L$, where $\epsilon_0$ is the static dielectric constant of
the material, $q$ is the net charge of the vacancy, and $\alpha$ is
the appropriate Madelung constant. Using this formula, we can subtract
off the leading finite-size corrections, and greatly accelerate the
convergence with respect to supercell size. In applying this
correction procedure in the present work, we take $\epsilon_0$ for
bulk MgO from experiment~\cite{peckham67}.

The present DFT calculations are performed using the VASP
code~\cite{kresse96} on a wide range of periodic cell sizes, as
described above. These calculations allow us to determine the cell
size needed to obtain converged results. The calculations employ the
projector augmented wave method~\cite{blochl94,kresse99} for the
interactions between the valence electrons and the ions, and the local
density approximation for electronic exchange and correlation. In all
the DFT calculations, the entire system is fully relaxed, so that the
maximum force on any ionic core is less than $ 4 \times
10^{-4}$~eV~\AA$^{-1}$.

Detailed descriptions of VMC and DMC and of the {\sc casino} code used
for all the QMC calculations have been reported
elsewhere~\cite{foulkes01,needs04}.  Our trial wavefunctions
$\Psi_{\rm T}$ have the usual Slater-Jastrow form $\Psi_{\rm T} =
D^{\uparrow} D^{\downarrow} \exp ( J )$, where $D^\uparrow$ and
$D^\downarrow$ are Slater determinants of up- and down-spin
single-electron orbitals, and $\exp ( J )$ is the Jastrow factor
describing correlations between electrons. The function $J$ is a sum
of parameterised one- and two-body terms, the latter being designed to
satisfy the cusp conditions. The free parameters in $J$ are determined
by requiring that the variance of the local energy in VMC be as small
as possible. The many-body wavefunction represents explicitly only
valence electrons, whose interactions with the ionic cores are
described by pseudopotentials. The Hartree-Fock pseudopotentials used
here are the same as those used in our recent QMC work on bulk
MgO~\cite{alfe05}. The single particle orbitals have been taken from
DFT-LDA calculations with the same pseudopotentials using the {\sc
pwscf} code~\cite{pwscf}.  The basis set used for the representation
of the single-particle orbitals in QMC is the recently described
B-spline or blip-function basis~\cite{alfe04}, which is closely
related to plane waves, but is computationally far more efficient for
QMC. Since we cannot perform structural relaxations with QMC, the
relaxed ionic positions used in our QMC calculations on the defective
systems are taken from our DFT calculations.

In order to suppress statistical bias, QMC calculations need to be run
with a large population of ``walkers'', and this makes it efficient to
run them on large parallel machines.  The present calculations were
performed on the HPCx machine at Daresbury using between 128 and 320
processors, with a target number of 640 walkers.

In Fig. 1 we display the value of Schottky energy calculated using DFT
on various cell sizes, going up to 1024 atoms~\cite{problems}. The
results include the Coulomb correction mentioned
above~\cite{leslie85}. Two sets of calculations are reported in the
figure: one performed by sampling the Brillouin Zone (BZ) at the
$\Gamma$-point only, and the second using a $2 \times 2 \times
2$~Monkhorst-Pack~\cite{monkhorst76} grid. The two sets of results
become essentially indistinguishable for cells containing 128 atoms or
more and converge very quickly to the value of 6.76 eV. The error in
the Schottky energy obtained from calculations on cells containing
only 54 atoms is somewhat less than 0.2~eV, but we note that
cancellation of BZ errors makes the results obtained with 54 atoms and
$\Gamma$-point only sampling already converged to within $\sim
0.07$~eV.

DMC calculations have been performed on cells containing 54 atoms,
using a time step of 0.005~a.u. for over 50,000 steps. With this
length of simulations, total energies for the perfect MgO crystal and
the crystals with one Mg$^{2+}$ or one O$^{2-}$ vacancy were obtained
with statistical errors of $\sim 0.23$~eV, $\sim 0.32$~eV and $\sim
0.27$~eV respectively. The value of the Schottky energy with the error
bar obtained by combining the errors on the various components is $7.5
\pm 0.53$~eV. For completeness, we also report the value of the
Schottky energy of 6.99~eV obtained using DFT-LDA with the same
pseudopotentials used in the DMC calculations. Although there are
accurate experimental data for the migration energies of
cation and anion vacancies in MgO, the Schottky energy
itself is experimentally uncertain, with measured values
spanning the range $4 - 7$~eV~\cite{mackrodt82}. Our QMC
value is consistent with this, and also with earlier predictions
for $E_{\rm S}$ from both DFT~\cite{devita92} and calculations
based on interaction models~\cite{catlow76,mackrodt79}, all
of which give values in the range $6.5 - 7.5$~eV.

These results demonstrate the technical feasibility of using high
precision QMC calculations to study the energetics of defects in oxide
materials. This is encouraging, because it suggests the possibility of
using QMC to calculate the formation, association and migration
energies of other kinds of defects, including impurities in oxide
materials. In the present work, the QMC result
for the Schottky energy agrees with the DFT predictions to within the
QMC statistical error, but this is expected because DFT is known to
give a good description of most properties of MgO. However, this will
not be true of strongly correlated materials such
as transition metal oxides, for which DFT
predictions are often poor. We are currently attempting to
extend our QMC calculations to the important oxide FeO.

A technical difficulty that is clear from the present work is that
very long QMC runs are needed to reduce the statistical error on the
defect energy to an acceptable level. This difficulty can be mitigated
by improving the quality of the trial wavefunction. In the present
case, we might do this by allowing the Jastrow factor to be different
in the vacancy region. The use of recently developed methods for
improving the scaling of the computer effort with system
size~\cite{williamson01,manten03,alfe04b} (so-called $O(N)$ QMC) will
also help in future work. But the fundamental problem is that there is
no cancellation of statistical noise on the large energies that are
subtracted. Correlated sampling may perhaps help with this, but the
solution may lie also in the close connection between $O(N)$ and
embedding~\cite{bowler02} that has already explored within
tight-binding and DFT methods.

DA acknowledges support from the Royal Society and from the Leverhulme
Trust. Allocation of computer time at the HPCx national service was
provided by the Materials Chemistry Consortium (EPSRC grant GR/S13422)
and at the CSAR national service by the Mineral Consortium.

\begin{figure}
\psfig{figure=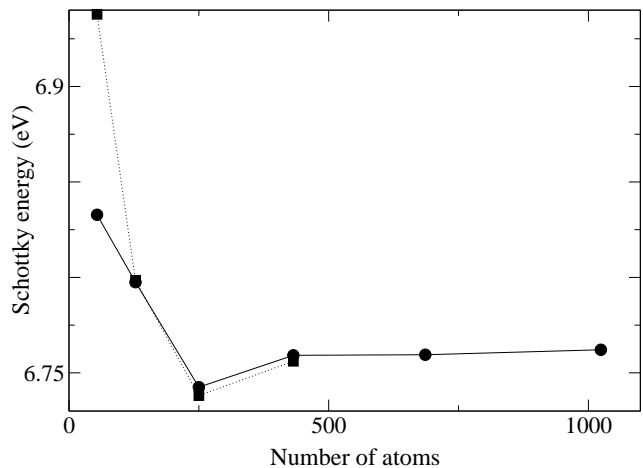,height=3.0in,angle=-90}
\caption{Schottky vacancy formation energy (eV) in MgO calculated in the
local density approximation of DFT for repeating cells containing
different number of atoms. Filled circles connected by solid lines
(filled squares conected by dotted lines) show results obtained with
$\Gamma$-point ($2 \times 2 \times 2$~Monkhorst-Pack) Brillouin zone
sampling. }
\end{figure}


\begin{thebibliography}{99}
\bibitem{problems} We note in passing a technical problem that we
encountered in the DFT calculations. The oxygen vacancy induces a gap
state whose energy depends on the ionic positions. In some
circumstances, partial occupation of this state can hinder the search
for the ground state. However, for the relaxed system, the gap state
lies well above the top of the valence band, and becomes completely
unoccupied, so that the problem disappears.
\end{thebibliography}
\end{document}